\documentclass[useAMS,referee,usenatbib]{biom}

\usepackage{bm}
\usepackage{amsmath,amsopn,amssymb}

\usepackage{enumitem}
\usepackage{graphicx}
\usepackage{titlesec}
\usepackage{booktabs}
\usepackage{threeparttable}
\usepackage[colorlinks,citecolor=blue,urlcolor=blue,filecolor=blue]{hyperref}

\title[BISON]{BISON: Bi-clustering of spatial omics data with feature selection}

\author{Bencong Zhu $^{1,2}$, Alberto Cassese$^3$, Marina Vannucci$^4$, Michele Guindani$^5$, Qiwei Li$^{2, \ast}$ \\ 
$^{1}$Department of Statistics, The Chinese University of Hong Kong, Hong Kong SAR, China \\
$^{2}$Department of Mathematical Sciences, The University of Texas at Dallas, Richardson, Texas, U.S.A \\
$^{3}$Department of Statistics, Computer Science, Applications ``G. Parenti'', University of Florence, Florence, Italy \\
$^{4}$Department of Statistics, Rice University, TX, U.S.A. \\
$^{5}$Department of Biostatistics, University of California, Los Angeles, CA, U.S.A.
}

\begin{document}


\date{{\it Received October} 2007. {\it Revised February} 2008.  {\it
Accepted March} 2008.}



\pagerange{\pageref{firstpage}--\pageref{lastpage}} 
\volume{64}
\pubyear{2008}
\artmonth{December}


\doi{10.1111/j.1541-0420.2005.00454.x}


\label{firstpage}

\begin{abstract}
The advent of next-generation sequencing-based spatially resolved transcriptomics (SRT) techniques has reshaped genomic studies by enabling high-throughput gene expression profiling while preserving spatial and morphological context. Understanding gene functions and interactions in different spatial domains is crucial, as it can enhance our comprehension of biological mechanisms, such as cancer-immune interactions and cell differentiation in various regions. It is necessary to cluster tissue regions into distinct spatial domains and identify discriminating genes that elucidate the clustering result, referred to as spatial domain-specific discriminating genes (DGs). Existing methods for identifying these genes typically rely on a two-stage approach, which can lead to the phenomenon known as \textit{double-dipping}. To address the challenge, we propose a unified Bayesian latent block model that simultaneously detects a list of DGs contributing to spatial domain identification while clustering these DGs and spatial locations. The efficacy of our proposed method is validated through a series of simulation experiments, and its capability to identify DGs is demonstrated through applications to benchmark SRT datasets.\\
\end{abstract}



\maketitle

\section{Introduction}
Spatially resolved transcriptomics (SRT) technologies have been rapidly developed and widely used in biomedical research over the past years. These innovative technologies fall into two mainstreams: 1) imaging-based SRT platforms, including seqFISH \citep{lubeck2014single}, MERFISH \citep{chen2015spatially}, STARmap \citep{wang2018three}, etc.; and 2) next-generation sequencing (NGS)-based SRT platforms, such as spatial transcriptomics (ST) \citep{staahl2016visualization}, 10x Visium ST, high-definition ST \citep{vickovic2019high}, Slide-seq \citep{rodriques2019slide}, etc. The former are typically limited to hundreds of pre-selected genes, whereas the latter can reconstruct a transcriptome-wide spatial map covering expression levels of tens of thousands of protein-coding genes, providing a more comprehensive understanding. With these advancements, NGS-based SRT techniques have become pivotal in discovering novel insights in biomedical research. 

The rise of spatial transcriptomics has motivated the development of new statistical methods that handle the identification of {spatially variable genes} (SVGs), that is, genes with spatial patterns of expression variation across the tissue sample. Recently, \cite{yan2024categorization} have summarized $34$ state-of-the-art methods associated with SVG detection, categorizing
SVGs into three types: overall, cell type-specific, and spatial domain-marker SVGs. The overall SVGs are defined as the genes that exhibit non-random spatial expression patterns, whose representative detection methods include Trendsceek \citep{edsgard2018identification}, spatialDE \citep{svensson2018spatialde}, SPARK \citep{sun2020statistical}, BOOST-GP \citep{li2021bayesian}, BOOST-MI \citep{jiang2022bayesian}, BOOST-HMI \citep{yang2024bayesian}, BSP \citep{wang2023dimension}, etc. The cell type-specific SVGs are the genes that exhibit non-random spatial expression patterns within a cell type. The related methods include CTSV \citep{yu2022identification}, C-SIDE \citep{cable2022cell}, and spVC \citep{yu2024spvc}, ultilize both SRT data and external cell type annotations.  The spatial domain-marker SVGs are defined as the genes that exhibit significantly higher expression in a spatial domain compared to other domains. The methods for detecting spatial domain-marker SVGs first partition the tissue into multiple mutually exclusive spatial domains and then conduct hypothesis tests to evaluate differences in gene mean expression across these spatial domains. For example, SpaGCN \citep{hu2021spagcn} identifies spatial domains using a pre-trained graph convolutional network applied to SRT data and the paired histology image. Then, for each gene, it performs Wilcoxon rank-sum tests on normalized expression levels between each domain and the neighboring spots. DESpace \citep{cai2024despace} first implements existing spatial clustering methods, such as BayesSpace \citep{zhao2021spatial}, to identify spatial domains and then uses a generalized linear model based on negative binomial distribution to assess if the spatial domains significantly affect a gene’s expression, similar to iIMPACT \citep{jiang2023integrating}.

Despite the large amount of work aforementioned, challenges remain in detecting SVGs, particularly spatial domain-marker SVGs. Firstly, those heuristic two-step procedures for spatial domain-marker SVGs may accumulate estimation errors at each step, leading to an inflated false positive rate. This issue, known as \textit{double-dipping}, arises when the same dataset is used to define spatial clusters (e.g., cell types) and subsequently test for differential gene expression across those clusters \citep{neufeld2024inference, wang2024false}.
Second, some biologically relevant genes may exhibit high expression only within small regions of interest and can be overlooked by methods that fail to account for such localized expression patterns \citep{yan2024categorization}. 
To address this, \cite{sottosanti2023co} proposed  SpaRTaCo, a Gaussian process-based latent block model  to partition gene expression profiles in SRT data into several blocks, thereby identifying highly expressed marker genes within each spatial domain. However, by relying on all gene features, including non-informative ones that lack heterogeneity across spatial domains, this method could introduce noise and complicates the spatial domain identification process. Recent studies \citep{li2024interpretable,zhu2023bayesian} suggest that eliminating non-informative genes can substantially improve the accuracy of spatial domain identification and enhance downstream biological analyses.

\begin{figure}
\begin{center}
\includegraphics[width = \textwidth]{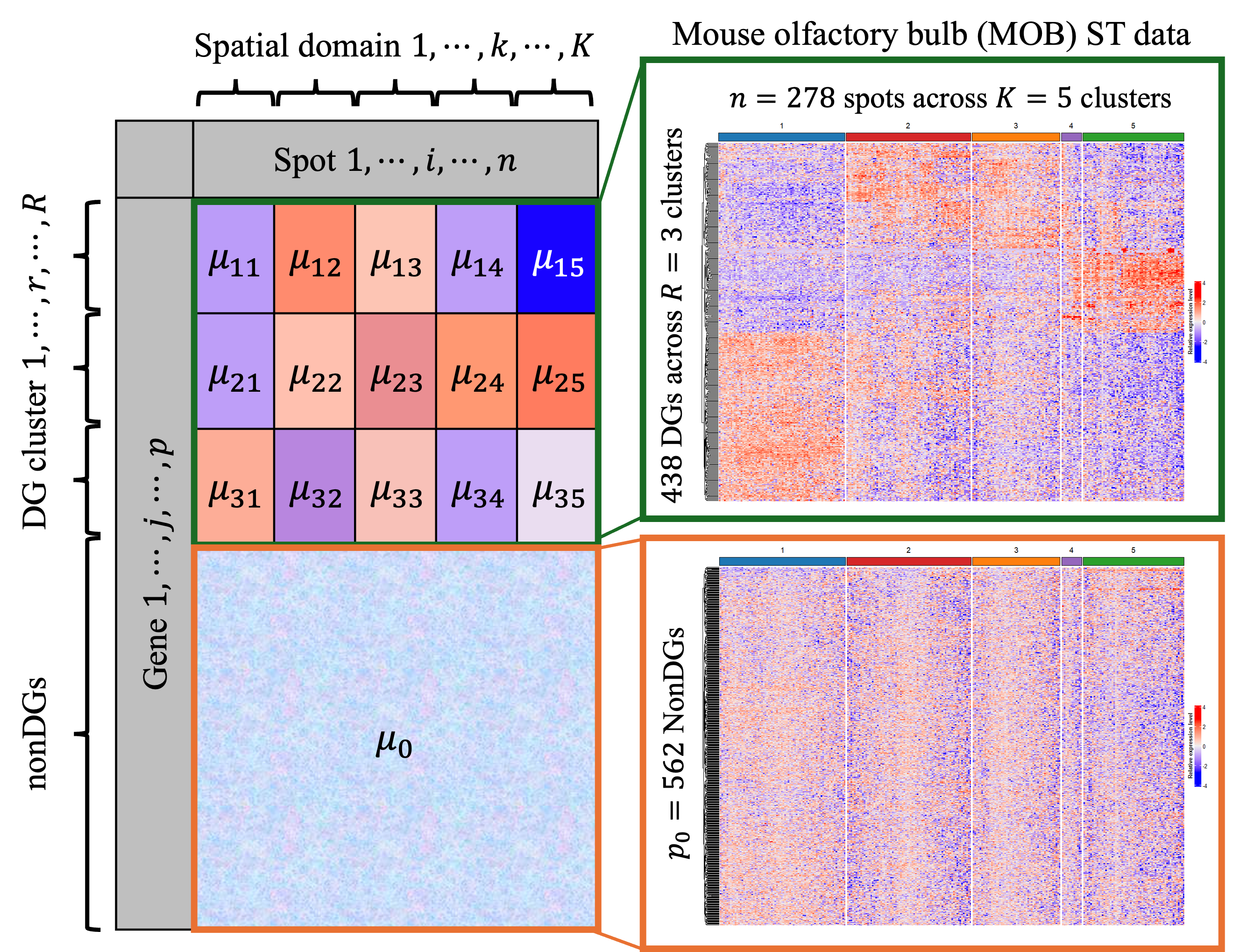}
\end{center}
\caption{An illustration of the proposed latent block model, BISON, with the null set parameterized by $\mu_0$, along with the resulting blocks from the MOB ST data analyzed in Section \ref{case_mob}.}
 \label{fig:schemetic}
\end{figure}

In response, we develop a unified Bayesian latent block model for bi-clustering of spatial omics data (BISON) with feature selection. BISON simultaneously identifies informative genes that contribute to spatial domain identification and clusters both these genes and spatial spots, as illustrated in Figure~\ref{fig:schemetic}. Our key contributions are as follows. First, BISON employs a multivariate count-generating process based on a Poisson model to directly model SRT count data, eliminating the need for \textit{ad hoc} data normalization methods. Second, BISON incorporates a feature selection strategy to generate a list of spatial domain-specific discriminating genes (DGs), enabling a lower-dimensional yet biologically interpretable representation of SRT data. Third, BISON uses a Markov random field (MRF) prior to account for the geospatial structure of SRT data, facilitating the mapping of contiguous domains. Lastly, we introduce a modified integrated complete likelihood criterion to determine the number of gene groups and spatial domains. The effectiveness of BISON is demonstrated through extensive simulation studies. In applications to mouse olfactory bulb ST data and human breast cancer 10x Visium data, BISON outperforms SpaRTaCo and other competing methods. Moreover, the DG groups identified by BISON are highly expressed within distinct spatial domains, indicating that each group represents spatial domain-marker genes specific to different spatial domains.

The remainder of the paper is organized as follows: Section \ref{sec:methods} presents the detailed formulation of the proposed bi-clustering framework with feature selection, along with posterior inference and model selection.
In Sections \ref{sec:sim} and \ref{sec:app}, we evaluate the proposed model using simulated data and two real SRT datasets. Section \ref{sec:con} concludes the paper with a brief discussion about the limitations and potential avenues for future research.

%
%

\section{Methods}
\label{sec:methods}
NGS-based SRT platforms, such as ST and the enhanced 10x Visium platforms, measure genome-wide expression levels encompassing over ten thousand genes across thousands of spatial locations referred to as `spots' on a tissue section. The molecular profile is represented by a $p \times n$ count matrix $\bm{Y}_{p \times n}$, with each entry $y_{ji} \in \mathbb{N}$ being the read count for gene $j$ observed in spot $i$. Here, we use $i=1,\ldots,n$ and $j=1,\ldots,p$ to index spots and genes, respectively. The corresponding geospatial profile is depicted by an $n \times 2$ matrix $\bm{T}_{n \times 2}$, where the $i$-th row $\bm{t}_{i\cdot}=(t_{i1},t_{i2})\in\mathbb{R}^2$ gives the x and y-coordinates of the $i$-th spot's center. Notably, the $n$ spots are approximately arrayed on two-dimensional square or triangular lattices, with each interior spot surrounded by four or six adjacent spots in the ST and 10x Visium platforms, as shown in Figure S8. This spatial configuration allows us to alternatively define the geospatial profile by an $n\times n$ binary adjacency matrix $\bm{E}$, where each entry $e_{ii'}=1$ if spot $i$ and $i'$ are neighbors and $e_{ii'}=0$ otherwise. 
Note that all diagonal entries in $\bm{E}$ are set to zero by default, i.e., $e_{ii}=0,\forall i$. In the following subsections, we provide the detailed formulation for BISON, with its hierarchical model graphically represented in Figure S1.

\subsection{Model formulation}
Within DGs, we assume there are $K$ spatial domains for spots and $R$ gene groups, resulting in $R\times K$ latent blocks. We introduce the latent vectors $\bm z = (z_{1}, \ldots,z_{n})^{\top}$ and $\bm \rho = (\rho_{1}, \ldots, \rho_{p})^{\top}$ to denote the latent cluster membership of columns (i.e., spots) and rows (i.e., genes), respectively. Specifically, we define $\mathcal{C}_{k} = \{i: z_{i} = k\}$ as the $k$-th column cluster and $\mathcal{D}_{r} = \{j: \rho_{j} = r\}$ as the $r$-th row cluster. The subset $\mathbf{Y}^{rk} = (y_{ji})_{j\in \mathcal{D}_{r}, i \in \mathcal{C}_{k}}$ denotes the observations in the $rk$-th  block. For each entry $y_{ji} \in \mathbf{Y}^{rk}$, we model the count using a Poisson distribution as follows
\begin{equation}
y_{ji}|z_{i} = k, \rho_{j} = r \sim  \text{Poi}(s_{i}g_{j}\mu_{rk}). \label{eq:lbm_model}
\end{equation}
We decompose the expected value of the Poisson distribution into a product of the scaling factors $s_{i} \in \mathbb{R}^{+}$ and $g_{j} \in \mathbb{R}^{+}$, which adjust for spot-specific effects and gene-specific effect, respectively, and the normalized gene expression level $\mu_{rk} \in \mathbb{R}^{+}$. Such a multiplicative formulation of Poisson means is typical in both the frequentist \citep{10.1214/11-AOAS493, li2012normalization, cameron2013regression} and the Bayesian literature \citep{banerjee2003hierarchical, airoldi2016improving,li2017bayesian} to accommodate latent heterogeneity and over-dispersion in count data. A simple and practical approach is to set each size factor $s_{i}$ proportional to the total sum of counts in the corresponding spot, i.e., $s_{i} \propto \sum_{j}y_{ij}$ \citep{sun2020statistical}, with the constraint $\sum_{i=1}^{n} s_{i} = 1$ to ensure identifiability \citep{li2021bayesian}. This leads to $s_{i} = \sum_{j=1}^{p}y_{ji} / (\sum_{i=1}^{n} \sum_{j=1}^{p}y_{ji})$. For the gene-specific effect $g_{j}$, we adopt $g_{j} = \sum_{i}y_{ji}$ as suggested by \cite{10.1214/11-AOAS493}. In the simulation study of Section \ref{sec:sim}, we  validate the effectiveness of the plug-in estimator, by comparing the estimated values $\hat{s}_{i}$ and $\hat{g}_{j}$ with the true $s_{i}$ and $g_{j}$. As shown in Figure S4, the estimation values are positively correlated with true values. For the prior of $\mu_{rk}$, we consider the conjugate prior $\mu_{rk} \sim \text{Ga}(\alpha_{\mu}, \beta_{\mu})$.

As suggested by \cite{zhu2023bayesian}, numerous non-discriminating genes (nonDGs) across spatial domains contribute minimal information for clustering the spots (i.e., columns). Including such non-informative genes may not only complicate the clustering process but also hinder the identification of true column clusters \citep{tadesse2005bayesian}. Here we define nonDG as genes whose expression has no heterogeneity across all spots after adjusting spot and gene-specific effects, implying that the normalized expression level parameters in Equation \ref{eq:lbm_model} is constant. To identify nonDGs, we introduce the null gene set $\mathcal{D}_{0} = \{j \in \left\{ 1,\ldots, p \right\} : \rho_{j} = 0\}$, whereby we can consider a total of $R+1$ gene clusters for the whole set of genes, for notational simplicity.  Thus, conditioning on $\rho_{j} = 0$,  the distribution of observations in the null set $\mathcal{D}_{0}$ can be expressed by a Poisson model, 
\begin{equation}
y_{ji} | \rho_{j} = 0 \sim \text{Poi}(s_{i}g_{j}\mu_{0}),  \label{eq:model0}
\end{equation}
where we consider the conjugate prior $\mu_{0} \sim \text{Ga}(\alpha_{0}, \beta_{0})$. 

\subsection{Priors on gene and spot cluster memberships}
Let $p_{0} < p$ indicate the total number of nonDGs and, correspondingly, let $p - p_{0}$ be the number of DGs. Following \citep{sivaganesan2011bayesian,xu2013nonparametric}, we propose a zero-inflated P\'{o}lya urn prior for $\bm \rho = (\rho_{1}, \ldots, \rho_{p})^{\top}$: 
\begin{equation}
P(\bm \rho) = \pi_{0}^{p_{0}}(1-\pi_{0})^{p-p_{0}}\frac{\gamma^{R}\prod_{r=1}^{R}\Gamma(p_{r})}{\prod_{j=1}^{p-p_0}(\gamma + j -1)}
\end{equation}
where $p_{r} = |\mathcal{D}_{r}|$ for $r \in \{1, \ldots, R\}$ is the cardinality of gene cluster $r$ and $\gamma$ is the total mass parameter of the P\'{o}lya urn scheme. Under this model, $P(\rho_{j} = 0) = \pi_{0}$, i.e. gene $j$ is a nonDG with probability $\pi_{0}$. When $\rho_{j} \neq 0$, gene $j$ is assigned to an existing gene cluster $r$ with probability $p_{r}/(p - p_{0})$.
Finally, we consider a conjugate Beta prior for $\pi_{0}$ by choosing $\pi_{0} \sim \text{Be}(\alpha_{\pi}, \beta_{\pi})$. 

For the prior distribution of the spot cluster membership vector $\bm z$, we adopt a Markov random field (MRF) model to incorporate available spatial information. In the context of NGS-based SRT data, several statistical models use a similar approach to enhance the spatial coherence of neighboring spots \citep[see, e.g.,][]{zhu2018identification, liu2022joint, jiang2023integrating, li2024interpretable}. Under this framework, the conditional distribution of each $z_i$ can be expressed as
\begin{equation}
P(z_i=k|\bm z_{-i}) \propto \exp\left\{ b_k + h \sum_{i'=1}^n e_{ii'}\text{I}(z_{i'}=k)\right\},
\end{equation}
where $b_k$ and $h$ are hyperparameters to be chosen and $\bm{z}_{-i}$ denotes the vector of $\bm z$ excluding the $i$-th element. Here $b_k$ controls the prior abundance of each cluster, and $h$ controls the strength of spatial dependence. We can also write the joint MRF prior on $\bm{z}$ as
\begin{equation}\label{eq:mrf}
    P(\bm{z}) \propto \exp\left\{\sum_{k=1}^K  b_{k} \sum_{i=1}^{n}\text{I}(z_{i}= k) + h\sum_{i < i'}e_{ii'}\text{I}(z_{i} = z_{i'})\right\}.
\end{equation}

For ST and 10x Visium platforms, the binary adjacency matrix $\bm E$ is created based on their square and triangular lattices, respectively. Note that if a spot does not have any neighbors or $h = 0$, its prior distribution reduces to a multinomial prior with parameter $\bm{q} = (q_{1}, \ldots, q_{K})^\top$ where $q_{k} = \exp(b_k)/\sum_{k=1}^K \exp( b_k)$ is a multinomial logistic transformation of $ b_k$.

\subsection{Posterior inference}
Our study focuses on detecting gene clusters through the gene allocation vector $\bm{\rho}$ and identifying spatial domains through the cluster allocation vector $\bm{z}$. We aim to sample from the posterior distributions of the unknown parameters $\bm{\mu} = \left\{(\mu_{rk})_{R \times K}, \mu_{0}\right\}$,  $\bm{\rho}$, and $\bm{z}$. The joint posterior can be written as
\begin{equation}
\begin{aligned}
    P(& \bm{\mu}, \bm{\rho}, \bm{z} | \mathbf{Y})  \propto  P(\mathbf{Y}\mid \bm{\mu}, \bm{\rho}, \boldsymbol{z}) P(\bm{\mu} \mid \bm{\rho}, \bm{z}) P(\boldsymbol{\rho} ) P(\boldsymbol{z}) \\
    & \propto  \left\{\prod_{r= 1}^{R}\prod_{k=1}^{K}\prod_{i\in \mathcal{C}_{k}} \prod_{j \in \mathcal{D}_{r}}\text{Poi}(y_{ij}|s_{i}g_{j}\mu_{rk}) \text{Ga}(\mu_{rk}|\alpha_{\mu}, \beta_{\mu})  \right\}\\
    & \times \left\{\prod_{j \in \mathcal{D}_{0}}\prod_{i=1}^{n}\text{Poi}(y_{ij}|s_{i}g_{j}\mu_{0}) \text{Ga}(\mu_{0}|\alpha_0, \beta_0) \right\}  P(\bm{z})P(\bm{\rho}). 
\end{aligned}
\end{equation}

We developed a Markov Chain Monte Carlo (MCMC) algorithm with a collapsed Gibbs sampler to obtain the posterior samples of $\bm \rho$ and $\bm z$ iteratively by integrating out the model parameter $\bm \mu$. Detailed descriptions of the MCMC procedure are provided in Section S1. Posterior inference on the relevant parameters is achieved \textit{via} post-processing of the MCMC samples after discarding posterior samples in the burn-in period.

A practical challenge related to posterior inference is to summarize a distribution over random partitions. \cite{dahl2006model} addresses the problem by estimating the pairwise probability matrix (PPM). For spots, the $\text{PPM}^{spot}$ is an $n\times n$ symmetric matrix, which calculates the posterior pairwise probabilities of co-clustering; that is, the probability that spot $i$ and $i'$ are assigned to the same cluster: $\text{PPM}^{spot}_{i,i'}\approx\sum_{u=1}^U{I}(z_i^{(u)}=z_{i'}^{(u)})/U$, where $u = 1, \ldots, U$ represents the iterations after burn-in. A point estimate $\hat{\bm z}^{\text{PPM}}$ can then be obtained by obtaining one iteration closest to the PPM:	
\begin{equation}
\hat{\bm z}^{\text{PPM}}  =  \underset{\bm 1 \leq u \leq U}{\mathrm{argmin}}~\sum_{i<i'}
\big({I}({z}^{(u)}_i={z}^{(u)}_{i'})-\text{PPM}^{spot}_{ii'}\big)^2.
\end{equation}
Similarly, we can also obtain the posterior PPM estimate for $\bm \rho$.
The PPM estimate leverages information from all clusterings through the PPM, providing a comprehensive and representative summary of the clustering results.

\subsection{Model selection}
The number of column clusters $K$ and row clusters $R$ can be specified based on prior biological knowledge, if available, or, conditionally on the absence of nonDGs, determined using the Integrated Complete Likelihood criterion  \citep[ICL][]{biernacki2000assessing}. If $p_{0} = 0$, the ICL is given by
\begin{equation}
\begin{aligned}
\operatorname{ICL}(R, K) &= -\sum_{j=1}^{p}\log P(\bm y_{j}, \hat{\bm{z}}, \hat{\rho}_{j}; (\hat{\mu}_{rk})_{R \times K}) \\
& + \frac{K-1}{2} \log (n) +\frac{R-1}{2} \log (p) + \frac{K R \nu}{2} \log (n p)
\end{aligned}
\end{equation}
where $\nu$ is the number of parameters per block \citep{bouveyron2018functional}. In our model where the nonDGs are estimated, we propose a modified ICL (mICL) criterion which incorporates the contributions of both DGs and nonDGs. Specifically, we consider
\begin{equation}
\begin{aligned}
\operatorname{mICL}(R, K) &= -\sum_{j \notin \hat{\mathcal{D}}_{0}}\log{P(\bm y_{j}, \hat{\boldsymbol{z}}, \hat{\rho}_{j}; (\hat{\mu}_{rk})_{R \times K})} \\
& +\frac{K-1}{2} \log (n) +\frac{R-1}{2} \log (p-\hat{p}_{0})  \\
& +\frac{K R \nu}{2} \log \{n (p-\hat{p}_{0})\} \\
& -  \sum_{j \in \hat{\mathcal{D}}_{0}}\log P(\bm y_{j}; \hat{\mu}_{0}) + \frac{\hat{p_{0}}}{2} \log(n),
\end{aligned}
\end{equation}
where $\hat{\mathcal{D}}_{0} = \{j = 1, \ldots, p: \hat{\rho}_{j} = 0\}$ is the set of estimated nonDGs and $\hat{p}_{0} = |\hat{\mathcal{D}}_{0}|$ is the estimated nonDGs cardinality. The couple $(R, K)$ leading to the lowest mICL value is selected as the most appropriate choice for the data at hand. The details for the computation of mICL are provided in Section S3.

%
%

\section{Simulation study}\label{sec:sim}

\subsection{Data generation}
\label{subsec:data_generation}
 The spatial pattern in simulations is extracted from the MOB ST data, which contains 278 spots arranged on a square lattice, with the number of spatial domains set to $K = 4$. We evaluate the model's performance under various setting of the total number of genes and proportion of nonDGs. Specifically, the total number of genes is varied as $p \in \{500, 1000\}$ and the proportion of nonDGs is varied as $\pi_{0} \in \{0, 0.2, 0.4, 0.6, 0.8\}$. The number of gene clusters for discriminating genes is $R = 3$, leading $\boldsymbol{\mu}$ to be a $3 \times 4$ matrix. The observed expression of nonDGs is sampled from a Uniform distribution $\mu_{0} \sim \text{Unif}(2, 6)$. The parameter $\mu_{rk} = 4 + (k-1) \Delta + (l-1) \Delta$, for $k \in \{1, \ldots, K\}$ and $r \in \{1, \ldots, R\}$ with $\Delta \in \{0.5, 1, 1.5\}$. The parameter $\Delta$ controls the signal of the data generation process. On average, a smaller value of $\Delta$ leads to a weaker signal in the dataset. The size factors and gene-specific effects are independent and identically distributed, $s_{i} \sim \text{Unif}(0.5, 1.5)$ and $g_{j} \sim \text{Unif}(0.5, 1.5)$. Finally, the count matrix is generated as
\begin{equation*}
    y_{ij} | \mu_{rk},\mu_{0j},\rho_j, z_i=k\sim
	\begin{cases}
		\text{Poi}\left\{s_{i}g_{j}(\mu_{rk} + \epsilon_{ij})\right\}& \text{ if } \rho_j=r\\
		\text{Poi}\left\{s_{i}g_{j}(\mu_{0j}+ \epsilon_{ij})\right\} & \text{ if } \rho_j=0
	\end{cases},
\end{equation*}
where we add a uniform random noise $\epsilon_{ij} \sim \text{Unif}(-0.1, 0.1)$ on the mean expression level. In summary, the simulation scheme results in $2\times 5\times 3 = 30$ simulation scenarios. For each scenario, we generated $50$ replicated datasets. 

\subsection{Results}
We set the hyperparameters of BISON to ensure weakly informative priors. Specifically, the hyperparameters of the Gamma distribution on $\mu_{rk}$ are set as $\alpha_{\mu} = \beta_{\mu} = 1$, and similarly the hyperparameters of the Gamma distribution on $\mu_0$ are set as $\alpha_0 = \beta_0 = 1$. For the Beta distribution governing the probability of a nonDG, we set $\alpha_{\pi} = 1$ and $\beta_{\pi} = 1$, resulting in a uniform prior with an expected value of $0.5$. Regarding the hyperparameters of the MRF prior model, we use $b_{k} = 1$ and $h = 1$ following the recommendations of \cite{li2024interpretable}. The model was run for $10,000$ MCMC iterations, with the first $5,000$ iterations discarded as burn-in, leaving $U = 5,000$ posterior samples for analysis.

We compare the performance of BISON in clustering spots and genes against alternative state-of-the-art methods, including \textbf{SpaRTaCo} \citep{sottosanti2023co}, the biclustering algorithm \textbf{BC} and its sparse version \textbf{sparseBC} (with penalty $\lambda = 10$) in the \verb|R| package \verb|sparseBC| \citep{tan2014sparse}, and a naive two-directional \textbf{K-means} approach, i.e. applying the k-means algorithm separately to spots and genes. Detailed implementations of the competing methods are provided in Section S2.2.

To evaluate clustering performance based on spot membership $\bm z$ and gene membership $\bm \rho$ across various methods, we used the adjusted Rand index \citep[ARI,][]{santos2009use}. The ARI, which ranges from 0 to 1, measures the similarity between two partitions. We computed the ARI using the partition induced by each method under study and the partition used to generate the simulated data. Higher ARI values indicate more accurate clustering performance. 
A rigorous definition of ARI is provided in Section S2.1.

\begin{figure}
\begin{center}
\includegraphics[width = \columnwidth]{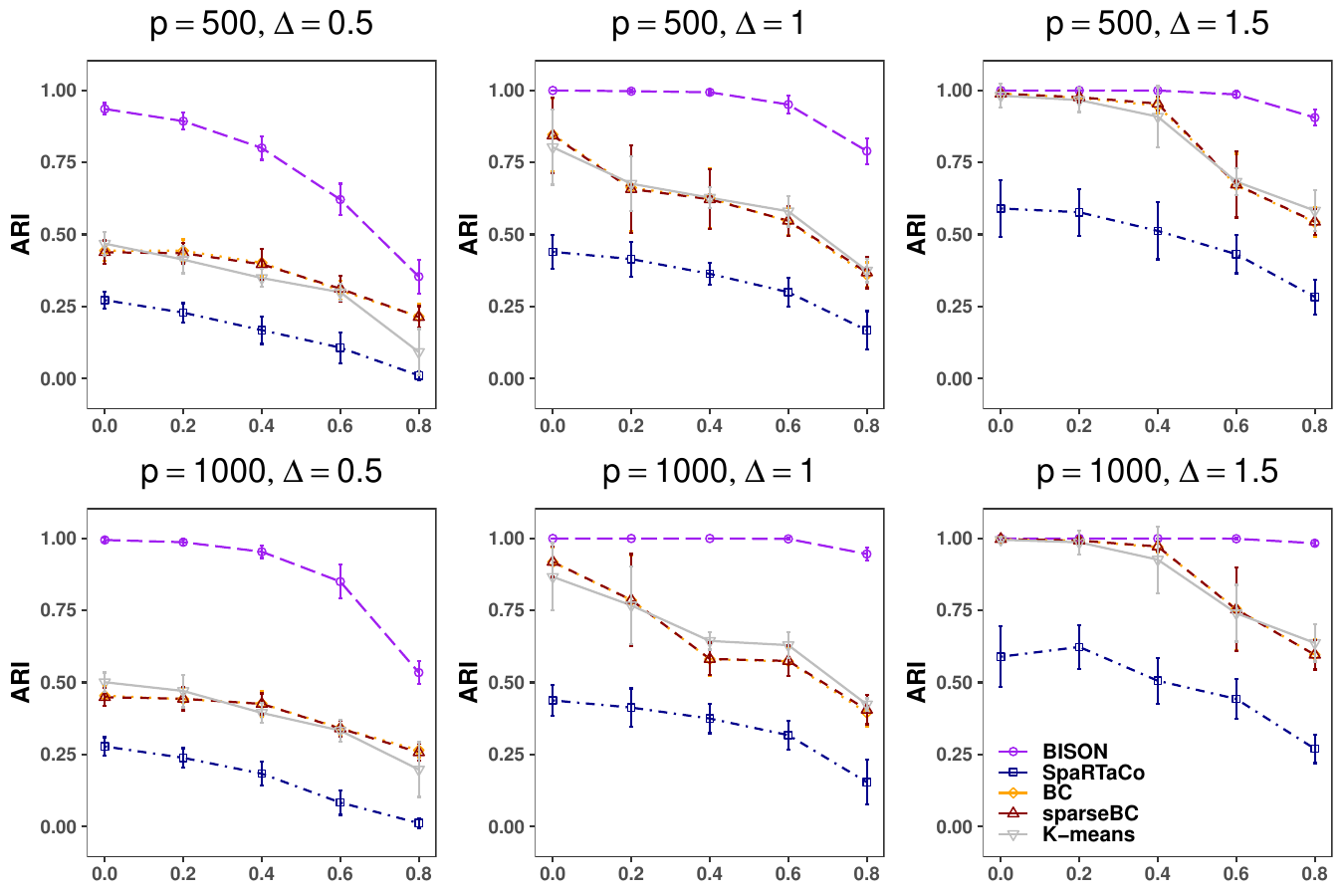}
\end{center}
\caption{Simulated results: Adjust rand index (ARI) for spots clustering against expected proportion of nonDGs $\pi_{0}$. Each subplot represents a specific combination of signal strength ($\Delta$) and the number of genes ($p$), as indicated on top of the subplot. For each scenario, the mean (point) and standard deviation (interval) of the ARI are computed across the 50 generated datasets.}
 \label{fig:ARI_col}
\end{figure}

Figure \ref{fig:ARI_col} illustrates the spots clustering performance of BISON across the various simulated scenarios, comparing it with the performance of competing methods in terms of ARI for spots clustering.
Overall, BISON consistently achieves the highest ARI for spots clustering score, indicating superior performance, especially when the signal is weak ($\Delta$ = 0.5). The performance of all methods deteriorates as the expected proportion of nonDGs ($\pi_{0}$) increases, suggesting that including nonDGs hinders the performance on spot clustering. Notably, BISON experiences a marked drop in performance when $\pi_0$ exceeds 0.6, whereas for the other methods, this decline occurs at lower values of $\pi_0$ and is less consistent. Interestingly, SpaRTaCo demonstrates the smoothest decline in performance as $\pi_0$ increases. Overall, the performance of the methods has similar patterns when comparing $p=500$ with $p=1,000$. When $\pi_0=0$, the scenario represents the absence of nonDGs.  BISON demonstrates superior performance under weak signals and maintains comparable performance under strong signals. This suggests that BISON is more robust compared to the alternative bi-clustering competing methods, in clustering spots. In terms of variability (standard deviation) of the results across the $50$ generated datasets per scenario, all the methods seem to have reasonable deviance.

\begin{figure}
\begin{center}
\includegraphics[width = \columnwidth]{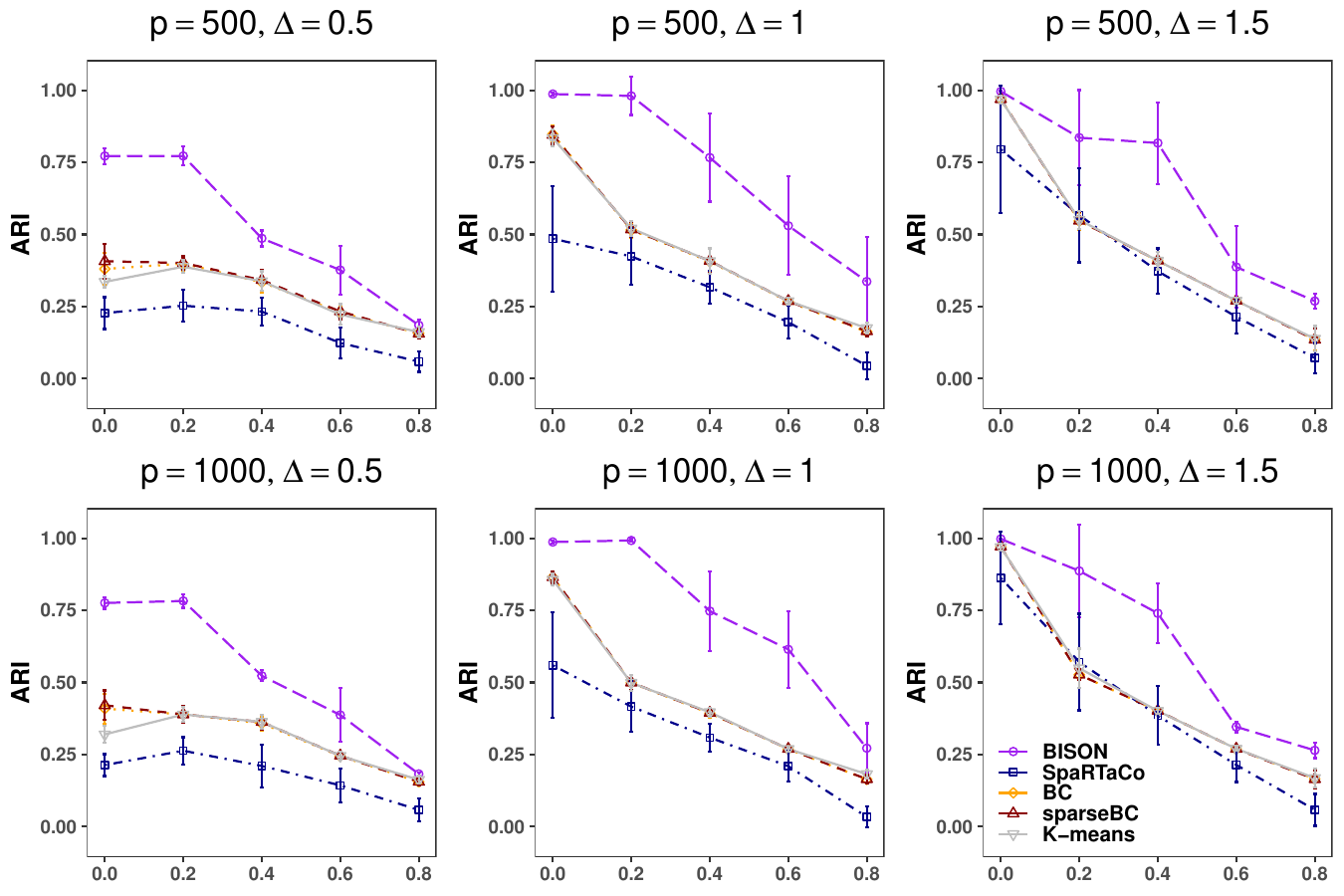}
\end{center}
\caption{Simulated results: The Adjust rand index (ARI) for gene clustering against expected proportion of nonDGs $\pi_{0}$. Each subplot represents a specific combination of signal strength ($\Delta$) and the number of genes ($p$), as indicated on top of the subplot. For each scenario, the mean (point) and standard deviation (interval) of the ARI are computed across the 50 generated datasets.}
 \label{fig:ARI_row}
\end{figure}

Figure \ref{fig:ARI_row} illustrates the clustering performance of BISON across the various simulated scenarios 
for genes clustering. 
Overall, BISON consistently achieves the highest ARI for genes clustering score, indicating superior performance. 
The performance of all methods also deteriorates dramatically as the expected proportion of nonDGs ($\pi_{0}$) increases. This may be attributed to the imbalance between the number of DGs and nonDGs. To further explore this phenomenon, we group all DGs into a single category and evaluate the performance in terms of DG identification. As shown in Figure S5, the specificity for detecting DGs declines sharply when $\pi_{0} > 0.6$, thus this only partially explains the observed results, as in most cases, for most methods the performance already deteriorates sharply when $\pi_0 > 0.2$. Overall, the performance of the methods has similar patterns when comparing $p=500$ with $p=1,000$. When comparing the results for different numbers of genes, the patterns and ARI levels are very similar for all methods. When $\pi_0=0$, the scenario represents the absence of nonDGs, creating conditions more favorable to the competing models. Nonetheless, BISON demonstrates superior performance under weak signals and maintains comparable performance under strong signals. 

Simulation results on clustering performance are based on setting the number of spot and gene clusters in the analysis to match the true values used to generate the simulated data. 
To further evaluate the mILC criterion's performance on the simulated datasets, we examined its results for BISON. Figures S6 and S7 show the performance in terms of the average estimated number of gene and spot clusters, respectively. 
Overall, the performance in terms of the number of spot clusters is robust. However, it shows limitations when the number of genes is small ($p=500$), the signal strength is weak ($\Delta=0.5$), and the expected proportion of nonDGs is high ($\pi_0=0.8$). Despite this, such scenarios are unlikely to pose significant issues in real datasets, as one would expect this proportion to be relatively small in many applications. For instance, the estimated proportion of nonDGs is $\approx 0.1$ in the HBC dataset and $\approx 0.5$ in the MOB dataset (see Section \ref{sec:app}). As for performance in terms of the number of clusters of genes, the results are similar and show only limitations when $\pi_0=0.8$. 

Finally, to investigate the robustness of the methods to model misspecification, we generate simulated data from a Negative Binomial distribution.
The Negative Binomial distribution introduces an additional dispersion parameter, $\psi_{j}$, which we simulate as $\psi_j\sim \mbox{Exp}(0.1)$. Consistent with the results presented earlier in this section, BISON continues to demonstrate the overall best performance, which are detailed in Section S2.3.

\section{Applications}\label{sec:app}

\subsection{Application to the MOB ST data}\label{case_mob}
We consider the MOB ST data, openly accessible through the Spatial Research Lab \footnote{https://www.spatialresearch.org/resources-published-datasets/doi-10-1126science-aaf2403/}. The preprocessed dataset comprises $n=278$ spots and $p=1,000$ highly variable genes. To motivate the application of BISON, we first performed an analysis of the MOB ST data where we implemented the clustering with feature selection method BNPSpace of \citep{zhu2023bayesian}. This method identifies spatial domains and detects the genes with heterogeneity among domains i.e. DGs. As shown in Figure \ref{fig:schemetic}, this analysis identified 5 domains and 438 DGs, which corresponds to about $44\%$ of the total genes considered. We applied a hierarchical clustering algorithm on the identified DGs, resulting in three distinct gene groups (Figures \ref{fig:schemetic} and S2). In contrast, applying the same algorithm to the nonDGs revealed no clear pattern (Figure \ref{fig:schemetic}. These findings illustrate that the expression of identified DGs is heterogeneous across domains, aligning with the definition of DGs. This highlights the need for a framework that can simultaneously discover these gene groups and spatial domains, essentially partitioning the expression matrix into non-overlapping rectangular blocks while excluding nonDGs whose presence could interfere with the accurate identification of the spatial domains.


We then applied BISON to analyze the preprocessed MOB data, using the same prior specifications and algorithm settings as in the simulation study. To determine the optimal number of gene clusters $R$ and spot clusters $K$, we computed the mICL for combinations of $K \in \{2, \ldots, 7\}$ and $R \in \{1, \ldots, 7\}$. Table S1  provides the top five mICL values, with the optimal model corresponding to $K = 4$ and $R = 3$. Interestingly, the spot clustering performance, measured by ARI, exhibits a negative correlation with mICL values. This relationship further supports the validity of the proposed mICL criterion, as lower values of mICL correspond to improved clustering results. To asses the MCMC convergence, we ran three independent MCMC chains with diverse initialization under the optimal choice of number of spot and gene clusters. Figure S10 presents the trace plots for these chains, which indicate satisfactory convergence. For posterior inference, we aggregated the outputs from all three chains. For the competing methods, we set the number of gene groups to $R+1$, allowing one group to capture the nonDG set. The number of spot clusters $K$ was kept consistent with the optimal choice used in BISON.

\begin{figure}
\begin{center}
\includegraphics[width = \columnwidth]{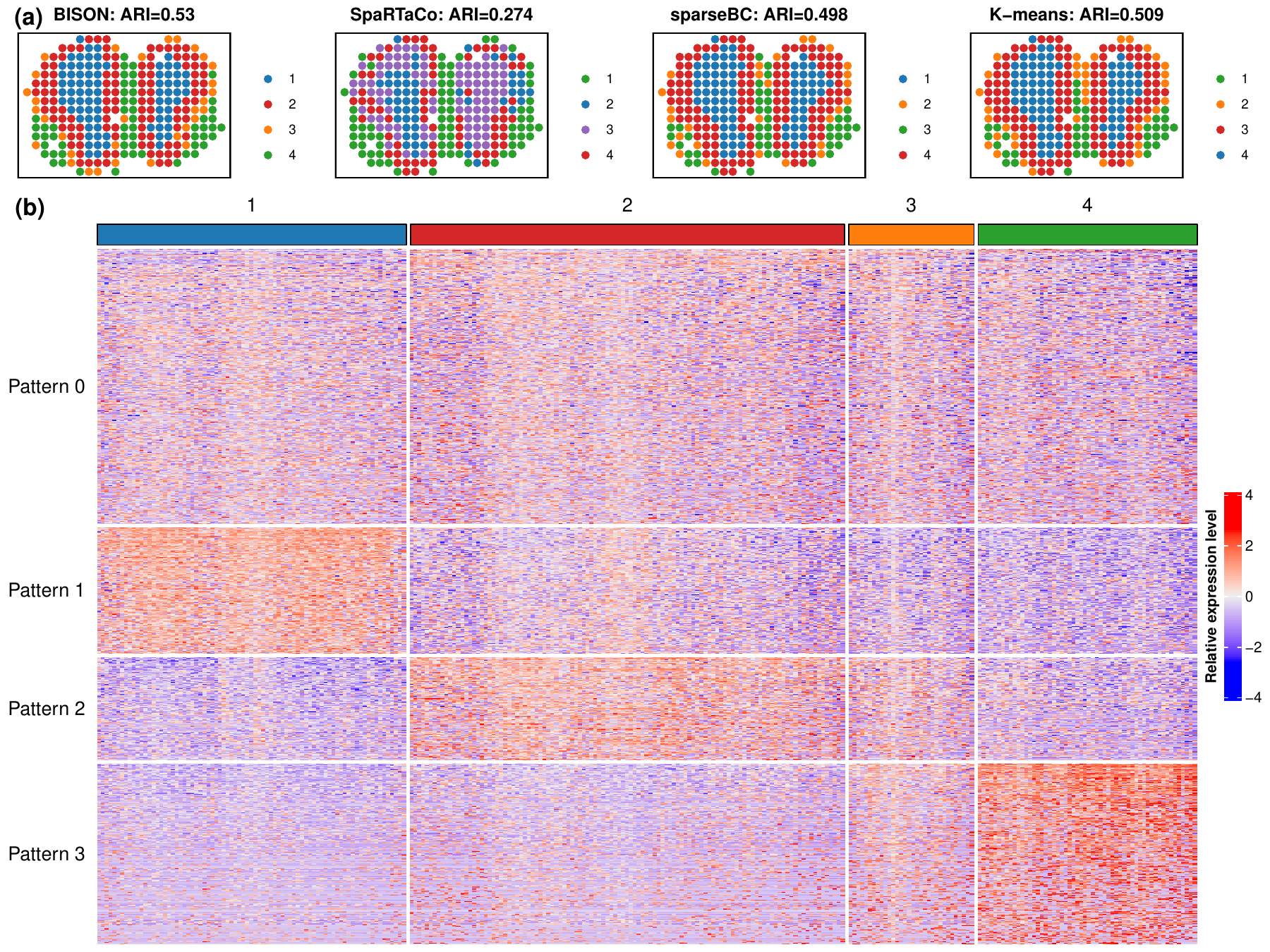}
\end{center}
\caption{MOB ST data: (a) The spatial domains identified by BISON and competing methods; (b) Heatmap of gene groups identified by BISON. Pattern 0 represents the nonDGs.}
 \label{fig:MOB_domain}
\end{figure}

\begin{figure}
\begin{center}
\includegraphics[width = \columnwidth]{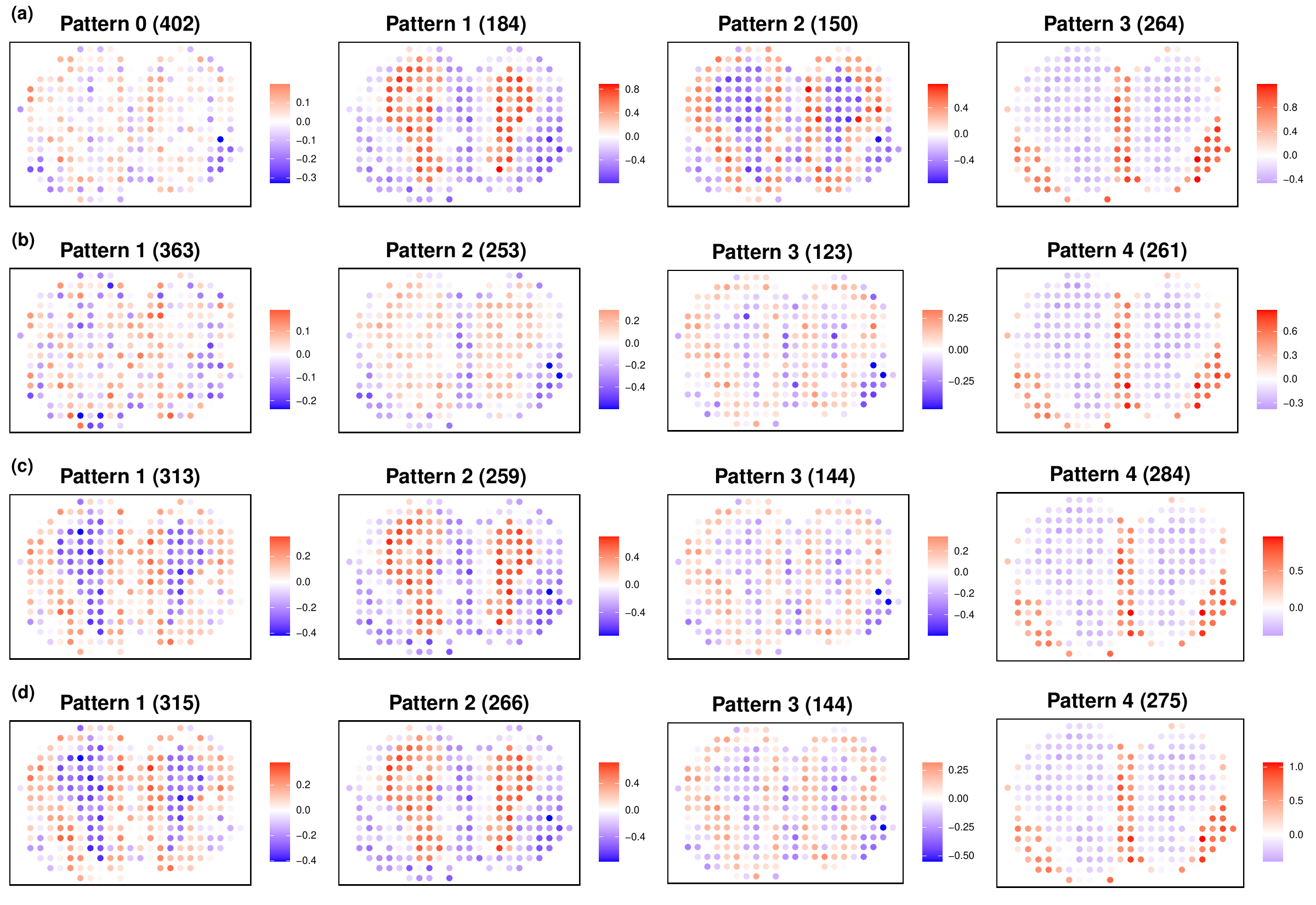}
\end{center}
\caption{MOB ST data: (a) The spatial expression patterns of gene groups identified by BISON; (b) The spatial expression patterns of gene groups identified by SpaRTaCo; (c) The spatial expression patterns of gene groups identified by sparseBC; (d) The spatial expression patterns of gene groups identified by K-means.}
 \label{fig:MOB_gene_group}
\end{figure}

The spots of MOB ST data were manually annotated by \cite{ma2022spatially} based on histology and serve as a benchmark to evaluate spot clustering performance. As shown in Figure \ref{fig:MOB_domain}(a), BISON achieved the highest concordance with the manual annotation, with the best ARI $=0.53$. The heatmap of gene expression in Figure \ref{fig:MOB_domain}(b) displays three DG groups alongside the nonDG group (Pattern 0). Specifically, the $184$ genes in the first gene group (Pattern 1) are highly expressed in the first spatial domain, corresponding to the inner layer of the tissue. The $160$ genes in the second gene group (Pattern 2) appear to be marker genes for the second and third spatial domains, corresponding to the middle layer of the tissue. Lastly, the $264$ genes in the third gene group (Pattern 3) are highly expressed in the last spatial domain, corresponding to the outer layer of the tissue. Figure \ref{fig:MOB_gene_group} depicts the spatial expression pattern for each gene group. Genes in Pattern 0 identified by BISON lack spatial information, whereas the genes in the three DG groups exhibit enriched spatial information across distinct spatial regions. Compared to the other bi-clustering methods, BISON discovered the most representative genes for the middle layer. These results align with the findings from the motivating example, further validating the effectiveness of BISON.

\subsection{Application to the HBC 10x Visium data}
We applied BISON to analyze an SRT dataset from a human breast cancer study, which includes 2,518 spots and 17,651 genes. The data set is publicly accessible on the 10x Genomics website \footnote{https://support.10xgenomics.com/spatial-gene-expression/datasets}. The gene expression was measured on a section of a human breast with invasive ductal carcinoma via the 10x Visium platform, along with partially annotated spatial domains from pathologists \citep{jiang2023integrating}. We keep the top $p = 1,000$ highly variable genes as the input of our analysis. Similarly as for the anlysis of the MOB dataset, we compute the mICL values for combinations of $K \in \{2, \ldots, 7\}$ and $R \in \{1, \ldots, 7\}$, and found $K = 5$ and $L = 4$ to be the optimal choice (Table S2).
We applied BISON, using the same prior specifications and algorithm settings as in the simulation study. Similarly, we ran three independent MCMC chains with diverse initialization under the optimal choice of the number of spot and gene clusters. Figure S11 presents the trace plots for these chains, which indicate satisfactory convergence. For posterior inference, we aggregated the outputs from all three chains. For the competing methods, we set the number of gene groups to $R+1$, allowing one group to capture the nonDG set. The number of spot clusters $K$ was kept consistent with the optimal choice used in BISON.

\begin{figure}
\begin{center}
\includegraphics[width = \columnwidth]{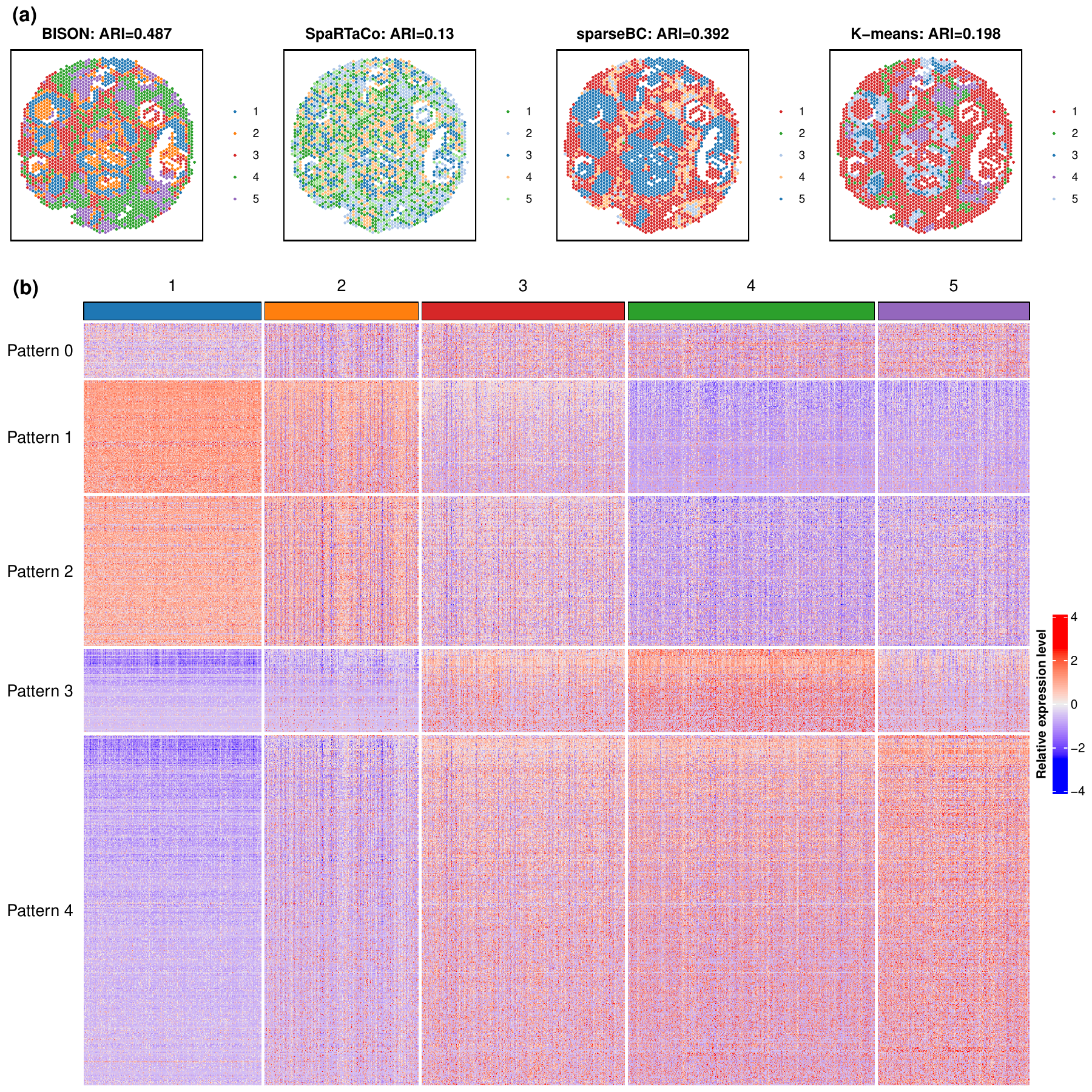}
\end{center}
\caption{HBC 10x Visium data: (a) The spatial domains identified by BISON and competing methods; (b) Heatmap of gene groups identified by BISON. Pattern 0 represents the nonDGs.}
 \label{fig:BC_domain}
\end{figure}

Figure \ref{fig:BC_domain}(a) shows the domains detected by the bi-clustering methods, BISON achieved the highest concordance with the manual annotation, with the best ARI $=0.487$. As for the gene groups identified by BISON, as shown in Figure \ref{fig:BC_domain}(b), the proportion of nonDGs in the HBC data is relatively small (Pattern 0 - 72 genes, $\hat{\pi}_{0} = 0.072$). Genes in the first gene group (Pattern 1 - 150 genes) and in the second gene group (Pattern 2 - 200 genes) are highly expressed in the first and second spatial domains. Their expression patterns are similar, but they are still identified as separate groups, since the mean expression levels are different, namely $\hat{\mu}_{11}/\hat{\mu}_{21} = 1.59$, and $\hat{\mu}_{12}/\hat{\mu}_{22} = 1.37$. The genes in the third gene group (Pattern 3 - $108$ genes) are highly expressed in the fourth spatial domain and further show a relatively high expression in the third domain.
The genes in the fourth gene group (Pattern 4 - $470$ genes) are highly expressed in the fifth spatial domain and further show a relatively high expression in the third and fourth domains. These two gene groups are quite similar to each other, but the third gene group is more representative of the fourth spatial domain, while the fourth gene group is more representative of the fifth spatial domain.

Last, we conducted gene ontology (GO) enrichment analysis for the identified spatial domain-marker genes using databases for annotation visualization and integrated discovery (DAVID) \citep{dennis2003david}. Annotation terms at a $0.05$ threshold applied to adjusted $p$-values were selected \citep{benjamini1995controlling}. As shown in Figure S13, genes in Pattern 1 and Pattern 2, which are highly expressed in tumor domains, are most enriched in the term, ``extracellular exosome'' (adjusted $p$-value $ < 2.7 \times 10^{-18}$). The exosomes have been reported to be the key players in cancer development provide theoretical supports for using exosomes to serve precise tumor treatment \citep{dai2020exosomes}. 
The number of significant GO terms for Pattern 3 and Pattern 4 is 59 and 239, respectively. Consequently, we present only the top 20 terms in Figure S14, where immune-related terms are identified, including ``immunoglobulin receptor binding'' ($p$-value $= 3.7 \times 10^{-10}$) and ``immune response'' ($p$-value $= 3.0 \times 10^{-13}$). 

\section{Conclusion}
\label{sec:con}
To identify spatial domains and spatial domain-marker genes simultaneously,  we have developed a unified Bayesian latent block model, namely BISON. 
The proposed modeling framework integrates feature selection and clustering of genes through a zero-inflated P\'{o}lya urn model,  leading to identifying a list of informative genes that contribute to the spatial domain identification while simultaneously clustering them. To obtain contiguous spatial domains, it
efficiently incorporates spatial information and achieves more robust and accurate spatial domains \textit{via}  a Markov Random Field (MRF) prior model. Finally, to determine the optimal number of gene groups and spatial domains, we propose a modified Integrated Completed Likelihood  criterion. 
 For efficient inference of model parameters, we developed a MCMC algorithm based on a collapsed Gibbs sampler. In our simulation study, BISON achieved superior performance in gene clustering and spatial domain identification under various proportions of nonDGs, different signal-to-noise levels, and gene cardinality. The model also achieved robust performance under model misspecification when simulated data were generated using a Negative Binomial model to produce the final counts. When applied to two real SRT datasets, genes in different discriminating gene groups were highly expressed in distinct spatial domains, indicating that the identified gene groups can serve as spatial domain-marker genes. Notably, BISON achieved the highest concordance with the available manual annotations, which serve as the benchmark.

Our model presents several limitations that merit future investigation.  
Firstly, the plug-in estimate of spot effect $s_{i}$ and gene effect $g_{j}$ may lead to biased estimation of the model parameters. These parameters can be estimated in a unified framework, as suggested by \cite{li2017bayesian}, albeit this would lead to a higher computational burden. Secondly, the question of whether to incorporate a zero-inflated component in SRT data modeling is still an active area of research. In this paper we have followed the suggestion of \cite{zhao2022modeling}, which states that a count model without a zero-inflation component is sufficient to fit the gene expression of SRT data. However, the model can be extended to a zero-inflated Poisson model or zero-inflated Negative Binomial model.

\bibliographystyle{biom} 
\bibliography{reference}

\label{lastpage}

\end{document}